\documentclass[A4,journal]{IEEEtran}
\usepackage{amsmath,amsfonts}
\usepackage{algorithmic}
\usepackage{algorithm}
\usepackage{array}
\usepackage[caption=false,font=normalsize,labelfont=sf,textfont=sf]{subfig}
\usepackage{textcomp}
\usepackage{stfloats}
\usepackage{url}
\usepackage{verbatim}
\usepackage{graphicx}
\usepackage{cite}
\begin{document}

\title{Smart City Drivers and Challenges in Urban-Mobility, Health-Care, and Interdependent Infrastructure Systems}

\author{Farid, Amro M., Alshareef, M., Badhesha, P.S., Boccaletti, C., Cacho, N.A.A., Carlier, C.-I., Corriveau, A., Khayal, I., Liner, B., Martins, Joberto S.B., Rahimi, F., Rossett, R., Schoonenberg, W.C.H., Stillwell, A., Wang, Y.

\thanks{Amro M. Farid is with Thayer School of Engineering, Dartmouth, USA; Muhannad Alshareef is with Al-Taqa Engineering, Pakistan; Parupkar Singh Badhesha is with Institute of Engineers, India; Chiara Boccaletti is with Sapienza University of Rome, Italy; Nelio Cacho is with Federal University of Rio Grande do Norte, Brazil; Claire-Isabelle Carlier is with Brookfield Renewable Partners LP, Canada; Amy Corriveau is with Emerging Business Development for CDM Smith, USA; Inas Khayal is with Dartmouth College, USA; Barry Liner is with Water Environment Federation, USA; Joberto S. B. Martins is with Salvador University (UNIFACS), Brazil; Farokh Rahimi is with Open Access Technology International, USA; Rosaldo Rossett is with University of Porto, Portugal; Wester C.H. Schoonenberg is with Thayer School of Engineering, USA; Ashlynn Stillwell is with University of Illinois, USA; Yinhai Wang is with University of Tokyo, Japan.}}

%\author{IEEE Publication Technology,~\IEEEmembership{Staff,~IEEE,}
        % <-this % stops a space
%\thanks{This paper was produced by the IEEE Publication Technology Group. They are in Piscataway, NJ.}% <-this % stops a space
%\thanks{Manuscript received April 19, 2021; revised August 16, 2021.}}

\markboth{IEEE Potentials,~Vol.~40, No.~1, Jan.-Feb.~2021}%
{Shell \MakeLowercase{\textit{et al.}}: A Sample Article Using IEEEtran.cls for IEEE Journals}


\maketitle

\begin{abstract}
At the turn of the 21st century, urban development has experienced a paradigm shift so that the quest for smarter cities has become a priority agenda, with the direct participation of industry, policymakers, practitioners, and the scientific community alike. The 2008 financial crisis, the exodus from rural areas, and the densification of urban centers coupled with environmental and sustainability concerns have posed enormous challenges to municipalities all over the globe. The United Nations predicts that the world population will reach 9.8 billion by 2050, a growth of 2.1 billion from the 2018 level. Almost all of this population growth will occur in urban areas and, consequently, stress already overloaded transportation systems. 
\end{abstract}


\section{Introduction}
At the turn of the 21st century, urban development has experienced a paradigm shift so that the quest for smarter cities has become a priority agenda, with the direct participation of industry, policymakers, practitioners, and the scientific community alike. The 2008 financial crisis, the exodus from rural areas, and the densification of urban centers coupled with environmental and sustainability concerns have posed enormous challenges to municipalities all over the globe. The United Nations predicts that the world population will reach 9.8 billion by 2050, a growth of 2.1 billion from the 2018 level. Almost all of this population growth will occur in urban areas and, consequently, stress already overloaded transportation systems.

To accommodate growing transportation needs, smart mobility solutions enabled by intelligent transportation systems (ITSs) are needed. ITSs have revolutionized both urban and rural mobility over the past four decades. Most of the recent progress stems from early initiatives in intelligent vehicle–highway systems in the United States, transport telematics in Europe, and congestion management in Japan.

\begin{figure*}[!t]
\centering
\includegraphics[width=4in]{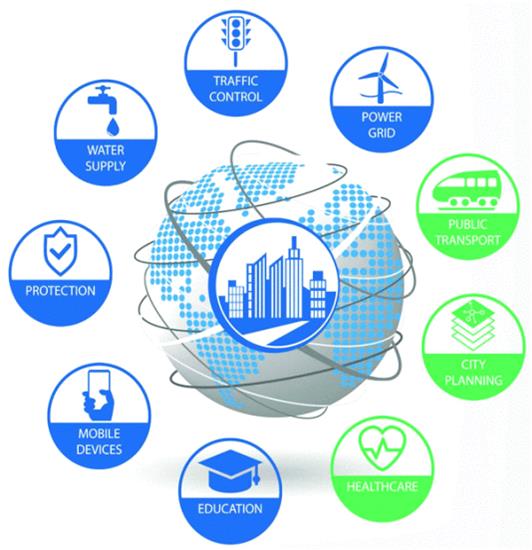}
\caption{Smart City}
\label{fig:SC}
\end{figure*}

Contemporary transportation heavily relies on information and communication technologies that support vehicle-to-vehicle (V2V) coordination, vehicle-to-infrastructure (V2I) interaction, and driverless autonomous cars operating in complex urban settings. With a focus on user experience, safety, efficiency, reliability, and accessibility, ITSs have achieved enormous progress and paved the way to a smart mobility paradigm, setting up the infrastructure and paving the road to a mobility paradigm shift integral to the realization of smart cities (Figure \ref{fig:SC}).

The role of the transportation system user is fundamentally evolving. Whereas ITSs and smart mobility are two concepts that are still used interchangeably, they are very much complementary. ITSs make users a central concern but, ultimately, treat them as passive entities. Meanwhile, in smart mobility, users become active players in the development and maintenance of transportation solutions. Consequently, there is a shift from intelligent systems that perform tasks correctly and efficiently to smart mobility, which accounts for subjective users’ perspectives and their personal interpretations of utility. This deep-seated consideration for human nature and preferences has proven to be the key ingredient for innovation in smart mobility systems.

\section{Drivers}

Although transportation systems face pressing needs in managing safety, congestion, reliability, cost, and sustainability, many of the most recent developments have been driven by technological innovation. These include the following:

\begin{itemize}
    \item Emerging vehicle technologies are in the process of enabling connected automated shared electric (CASE) vehicles as an end goal. CASE vehicles will change not only energy consumption patterns but also car ownership, land use, and mobility-as-a-service (MaaS).
    \item Beyond the vehicle, smart transportation infrastructure systems enable communication among vehicles, bicycles, pedestrians, and infrastructure. This enables optimal vehicle routing, efficient truck platooning, effective congestion mitigation, and quick incident response.
    \item The combination of crowdsourcing data, the Internet of Things (IoT), and big data analytics enables large-scale transportation system data collection and analysis for operational decision support and mobility service optimizations.
\end{itemize}

\section{Challenges}

Transportation is a fundamental need in a prosperous society and is directly correlated to a region’s per capita income. In the United States, traffic congestion cost US \$305 billion, and roadway crashes killed 37,133 people in 2017. Although improving transportation safety and congestion remain pressing smart mobility goals, significant technological and nontechnological challenges remain. Technological challenges include the following areas:

\begin{itemize}
    \item \textit{Sensing technology}: The detection of road conditions and surrounding objects is critical for safety in autonomous vehicle (AV) applications. The recent Uber and Waymo AV crashes show that there is still a long way to go to build reliable sensing systems that prevent the collision of AVs with surrounding objects in all weather and light conditions.
    \item \textit{High-fidelity mapping}: GPS based on satellite technology cannot offer the accuracy required by AVs. Researchers are working on high-resolution 3D maps to provide the needed accuracy. However, doing so opens further questions of how to communicate these data sets to AVs quickly and how to integrate dynamic roadway changes into the 3D maps in real-time.
    \item \textit{Vehicle and infrastructure coordination}: The U.S. Department of Transportation selected dedicated short-range communications (DSRC) as the IEEE Standard 802.11p protocol for standard V2V, V2I, and vehicle-to-pedestrian communication. Most mobile devices, however, do not support DSRC, and, consequently, implementation may be difficult and expensive. In contrast, 4G/LTE/5G technologies are easily accessible with existing personal devices. Connected vehicle and infrastructure solutions may choose to use 4G/LTE/5G technologies for user convenience at the potential expense of degraded cybersecurity.
\end{itemize}

Nontechnological challenges are equally important considerations in the implementation of smart mobility solutions. These include the following concerns:

\begin{itemize}
    \item \textit{Privacy}: MaaS works best when all service provider and user data are integrated into one single platform. The keeper of such data can wield tremendous power, and the natural question arises of whether a MaaS operator should be a for-profit, not-for-profit, or government entity. In the case of a for-profit entity, does this create the traditional problems associated with a monopoly, or can the benefits be rationalized as a “natural monopoly,” as in the case of many infrastructure systems? In the case of a government entity, many portions of society are wary of a \textit{1984}-esque “Big Brother” scenario. In the situation of a not-for-profit entity, is there sufficient incentive to produce efficient outcomes?
    \item \textit{Ethics}: In the event of a likely collision (which will ultimately occur), should an AV differentiate between a trajectory that minimizes damage to itself and its riders or damage of other vehicles? Human drivers, for example, have been shown to instinctively react in such a way as to protect the driver-side occupants at the expense of passenger-side riders.
    \item \textit{Social justice and equity}: Given that AVs are likely to be relatively higher cost than conventional vehicles, does this disparity create a systemic disadvantage to lower-income people?
    \item \textit{Civil and criminal justice}: How will law enforcement and insurance companies identify responsible parties and assess the associated penalties in the event of traffic violations and crashes? How will these entities take into consideration AVs with many different levels of autonomous driving?
\end{itemize}

Smart mobility solutions will bring huge benefits to smart cities and society as a whole. However, there are numerous challenges to address, and solving them requires the participation of stakeholders from a broad spectrum of fields, including law, social justice, technology, economics, and engineering.

\section{Smart health-care systems}

One of the primary missions of smart cities is to improve people’s quality of life, a critical component of which is their health. While health-related technology has significantly affected our ability to improve health, humanity is currently facing an unprecedented disease burden—so much so that, for the first time in history, our children’s generation is expected to have a shorter life span than our own. Furthermore, citizen disease burden manifests itself as significant health-care costs and the loss of economic opportunity for individuals and their caregivers.

Health is an emergent property of our health state over time, and smart cities have two specific roles in affecting it. First, cities develop—actively or passively—their inhabitants’ living environment, and, therefore, cities profoundly affect walkability, food environment and access, and exposure to environmental toxins. Second, cities have the ability to develop cyberphysical systems that allow citizens to access care when, how, and where they need it.

\subsection{Drivers}

Three primary drivers for addressing citizen health in smart cities are described.

\subsubsection{Growing financial incentives}

Annual health-care spending has grown faster relative to three relevant economic measures: the U.S. gross domestic product (GDP) growth, inflation, and population growth rates. It is now expected to grow to 20\% of the U.S. GDP by 2025. Such rising health-care cost projections have brought the issue of health-care delivery and prevention to the forefront of government budget discussions, from the federal level down to individual cities. Furthermore, many health-care insurance companies have designed plans that pass these growing expenses on to individuals, businesses, and cities.

In the meantime, the reimbursement mechanisms in these health-care plans are also being changed. Typically, health-care providers have been reimbursed on a “fee-for-service” basis. Such a mechanism incentivizes the provision of service without any focus on either the prevention of illness or health outcomes aftercare. An alternative reimbursement mechanism is based on “value for service.” It incentivizes payment based on the value measured in terms of health outcomes or improved prevention.

\subsubsection{Growing recognition of the importance of patient-centered personalized care}

The growing interest in the value-for-service reimbursement mechanism now requires a closer evaluation of the value of the health care that individuals receive. A growing body of research has shown that the variation in response to care exists for several reasons:

\begin{itemize}
    \item Variations in biological disposition;
    \item Differences in environmental factors present in a city; and
    \item Personal preferences, needs, and incentives.
\end{itemize}

Discerning these factors requires the ability to understand individuals more deeply. Consequently, measurement must transcend the traditional boundary of the health-care clinic and, instead, be embedded within the individual’s living environment. The ability to measure and transfer the associated information in near real-time is facilitated by the IoT, wearable sensors, and mobile health technologies.

\subsubsection{The ability to measure and deliver care facilitated by the IoT}

The IoT is a critical technical capability for measuring and delivering care where people live, work, and play. One of the primary limitations of classic health-care enterprises is that their health measurement and monitoring are restricted to noncontinuous and short-term schemes within health-care facilities. The IoT opens new opportunities for measuring health and delivering care outside the clinic so that it is continuous, event-driven, timed, or triggered at any location of choice—be it in a person’s home, community space, or work—in a personal or group setting. The ability to measure health and deliver care in a group setting introduces a social component to health-care delivery that has been shown to be very effective in recovery from many chronic illnesses.

\subsection{Challenges}

There are also several key challenges that need to be addressed to achieve improved individual health in smart cities. Three challenges are identified here.

\subsubsection{Lack of interconnectivity and interoperability of systems}

While the IoT has facilitated health sensing and care delivery to individuals’ living environments, such systems are typically developed independently and lack interconnectivity with other sensors or sensing platforms. This renders the independent systems noninteroperable and limits individuals’ ability to integrate multiple systems. The lack of interconnectivity and interoperability exists since there is no current incentive for makers of sensors to cooperate. At best, such device manufacturers seek to have their own proprietary self-interoperable platforms. However, stakeholders of value for service benefit from demanding interoperability from different sensors and systems so as to improve citizen health outcomes.

\subsubsection{Independent siloed health-care delivery systems}

Even though financial incentive drivers exist, it is difficult to capitalize on these drivers when health-care delivery systems are typically small, independent, and siloed in the types of care they provide. Consequently, individuals need to acquire care from multiple uncoordinated systems. Health-care delivery systems, while incentivized to attempt to treat individuals holistically and address their multiple needs, are neither structurally designed nor organized to deliver multiple care services. Large health-care systems, such as U.S. Veterans Affairs, Kaiser Permanente, and InterMountain Healthcare, have a structural advantage in providing patient-centered care. As a result, mergers and acquisitions are increasingly common in the health-care delivery industry.

\subsubsection{The rate of changing health-care information}

One important but often overlooked challenge is the ever-increasing quantity of health-care information. Consequently, health-care solutions and their associated technical and human resources must be developed in a way that facilitates flexibility and adaptation. The timely management and dissemination of health and clinical information pose further significant challenges.

\section{Smart interdependent infrastructure systems}

\subsection{Drivers}

While independent smart city infrastructure systems, such as energy, water, transportation, and health care, each have their own respective drivers and challenges, there is also a need to address the interdependencies of these infrastructures. The development of smart cities exists within a much larger trend of worldwide population growth and migration to urban areas.

These two trends, when combined, lead to the emergence of “megacities,” with populations of 10 million or more inhabitants. Furthermore, the majority of the megacities are located in coastal areas. Such cities are not only forced to accommodate an increasing number of people but are also exposed disproportionately to the effects of climate change, either as a result of extreme weather events or sea level rise. Consequently, their needs for citywide infrastructure resilience are often greater.

Infrastructure systems in cities are also increasingly interdependent as the urban population density increases. As smart cities become more crowded, systems theory recognizes that a large number of functional requirements imposed on a confined space (the city) inherently couple the elements of the systems. Consequently, infrastructures often have to serve multiple interdependent purposes—for example, an office building can be located on top of a metro station. An emergency in the office building may disrupt the metro service if the building needs to be evacuated.

The desire to create a more sustainable city also drives the integration of infrastructure systems. Transport is a major cause of pollution in cities. To reduce pollutants from transportation, cities encourage electrified public transportation and electric vehicles. Electrified transportation couples the electric power grid and transportation system, where each system relies on the other for its operation.

While such a scheme has the potential to enhance citywide sustainability, it also means that both infrastructures must function effectively in the event of extreme weather and disaster events. For example, during the electric power outages caused by Hurricane Sandy, it was difficult to evacuate New York City due to its reliance on the electrified subway system.

From a more technological perspective, the “future Internet” facilitates the development of “informatically integrated” interdependent infrastructure systems. It consists of the IoT, Internet of Services, and Internet of People. The future Internet acts as a platform to connect the smart city’s devices and sensors. This platform can be used for countless purposes, including information exchange, agent-based control, and large-scale data collection.

The impact of the “future Internet” is further magnified by the rapid expansion of big data analytics techniques, which provide new insights into the dynamics of, and interdependencies between, infrastructure systems by leveraging statistical patterns in historical data. Data are set to become increasingly ubiquitous as the future Internet evolves with its associated sensing capabilities. Nevertheless, big data analytics require “big theory” to understand the causal dynamics underlying these statistical patterns. Big data analytics support the identification of statistical correlations between infrastructure systems and, consequently, become the empirical basis for the development of an interdependent infrastructure systems theory.

\subsection{Challenges}

Despite these drivers, several challenges impede the successful implementation of interdependent smart city infrastructure systems. The elements of the interdependent infrastructure systems are heterogeneous and have dynamics distinct from those when each infrastructure is viewed independently.

For example, the electrified transportation system consists of the electric power grid and transportation system. The electrons on the electric power grid adhere to well-known physics, and the power grid operator has centralized control over the grid. On the other hand, the vehicles in the transportation system are controlled by humans and follow a distributed, agent-based behavior.

The integration of the electric power grid and transportation system, therefore, requires a deep understanding of the dynamics of each of the systems and the coupling between them. The emergence of big data facilitates the derivation of correlations between these systems; however, it does not provide a fundamental insight into the physics of the system as a whole.

The future Internet aims to connect all devices in a city on a single platform and provide them with the ability to communicate. The first challenge is that the presence, role, and value of private data on such a platform are unclear. The second challenge is that a standardized platform for communication has yet to emerge. Consequently, early-adopter cities are at risk of investing in multiple platforms that do not facilitate interoperable cross-platform communication. The uncertainty around the emergence of a dominant platform prevents cities from committing the required investments in a given technology.

Another challenge is that, as the physical infrastructure systems become increasingly interdependent, their respective institutions are compelled to collaborate. However, these institutions are not structured for effortless cross-infrastructure collaboration. Consequently, a consistent organizational framework needs to be implemented across multiple institutions that foster collaboration at a government level.

Another path, which by many would be viewed as a disruptive innovation, would be the development of new agencies with citywide jurisdiction over all urban infrastructure systems. Either way, before a city can truly commit to interdependent physical infrastructure, the organizational structure to support this integration needs to be in place.

These practical challenges hinder progress toward the development of interdependent infrastructure systems. However, some challenges have a more theoretical nature. Thus far, the scientific literature has generally studied interdependent infrastructures based on case studies. Though valuable for the administrators of that particular city, the generalizability of these case studies to other cities is limited.

Consequently, as each city becomes increasingly “smart,” it is forced to “reinvent the wheel” instead of learning from implementations elsewhere. Instead, the development of these solutions should be supported by best practices developed from the experiences of multiple cities. More specifically, a set of measures that values the heterogeneity of cities across the world should be defined. As a consequence of the previously defined drivers, measures for sustainability and resilience require specific attention.

Measures of sustainability use a variety of approaches. For products, it is common to use a lifecycle analysis. For energy supply systems, one of the measures of sustainability is the levelized carbon dioxide emissions. These measures, however, are not concerned with other measures of sustainability, such as the water footprint.

The measure of resilience for networks of infrastructure systems have based themselves predominantly on network theory. Independent infrastructure networks are abstracted as graphs, to which measures of resilience are applied. Network theory, however, is limited in its ability to represent heterogeneous networks, such as smart city interdependent infrastructure systems. One promising avenue for theoretical development is “heterofunctional graph theory,” which has been shown to not only model interdependent infrastructure networks but also provide analytical techniques for the measurement of systemwide resilience.

\section{Conclusion}
This article has served to highlight several domain-specific drivers and challenges within the broader smart city landscape. More specifically, it has viewed the increasing “intelligence” found in the energy, water, transportation, and health-care sectors. While each of these domains has its own specificities, it is clear that, from a smart city perspective, there are common themes. First, the IoT applies equally to all infrastructures. Each infrastructure system is at varying levels of development and deployment, but, ultimately, each infrastructure is robustly adopting the IoT paradigm.

Second, this increased adoption of IoT technology is leading to ever-greater distributed intelligence. At its heart, cities are encouraging empowered and engaged inhabitants that increasingly wish to play active roles in their quality of life and, ultimately, the infrastructure services that they receive and utilize in the city. Throughout all of the infrastructure systems mentioned, the end user is active and leverages information technology to gain a higher-quality infrastructure service. Furthermore, in sectors such as energy and transportation, the development of physical technologies like solar photovoltaics and electric vehicles serves only to accelerate this trend toward distributed intelligence.

Finally, all of the infrastructure systems discussed here are experiencing convergent interdependence. Each infrastructure system must now be viewed as part of an interdependent infrastructure system whole. In that regard, developments in information technologies must be supported by innovations in governance and theoretical developments to achieve systemwide benefits.

\section*{Read about it}

This is the second in a series of two articles that focus on challenges within a smart city landscape. The first article, “Smart City Drivers and Challenges in Energy and Water Systems” also appears in this issue of IEEE Potentials.

Additional reading: \cite{muhanji_eiot_2019}, \cite{schoonenberg_hetero-functional_2018},\cite{yih_handbook_2016},\cite{shaheen_future_2018},\cite{tsakalides_smart_2018},\cite{water_environment_federation_intelligent_2017}.

%\section{References Section}
%You can use a bibliography generated by BibTeX as a .bbl file.
% The IEEEtran BibTeX style support page is:
 

%\section{Simple References}
%You can manually copy in the resultant .bbl file and set second argument of $\backslash${\tt{begin}} to the number of references
% (used to reserve space for the reference number labels box).

\bibliographystyle{IEEEtran}
\typeout{}
\bibliography{ref.bib}

%\begin{thebibliography}{1}
%\bibliographystyle{IEEEtran}

%\bibitem{ref1}
%{\it{Mathematics Into Type}}. American Mathematical Society. [Online]. Available: https://www.ams.org/arc/styleguide/mit-2.pdf

%\end{thebibliography}

%\newpage

\vskip -2\baselineskip plus -1fil

\section{Biography}

\vskip -2\baselineskip plus -1fil

\begin{IEEEbiographynophoto} {Amro M. Farid} (amfarid@dartmouth.edu) is an associate professor of engineering at the Thayer School of Engineering at Dartmouth, Hanover, New Hampshire, USA. He leads the Laboratory for Intelligent Integrated Networks of Engineering Systems. He is a Senior Member of IEEE.
\end{IEEEbiographynophoto}

\vskip -2\baselineskip plus -1fil

\begin{IEEEbiographynophoto}
{Muhannad Alshareef} (muhannad3000@gmail.com) is a design and supervision engineer at Al-Taqa Engineering, Pakistan, for many electrical power, renewable energy, and building management system projects in the Middle East. He is a Member of IEEE.
\end{IEEEbiographynophoto}

\vskip -2\baselineskip plus -1fil

\begin{IEEEbiographynophoto}
{Parupkar Singh Badhesha} (psinghb2015@gmail.com) is a certified energy auditor and manager. He is a team leader for energy audits, automation, smart grids, renewable energy, and the construction and commissioning of hydro power plants with the state power utility in India. He is a Senior Member of IEEE and a fellow of the Institute of Engineers (India).
\end{IEEEbiographynophoto}

\vskip -2\baselineskip plus -1fil

\begin{IEEEbiographynophoto} {Chiara Boccaletti} (chiara.boccaletti@uniroma1.it) is an assistant professor at Sapienza University of Rome, Italy. She is a Senior Member of IEEE.
\end{IEEEbiographynophoto}

\vskip -2\baselineskip plus -1fil

\begin{IEEEbiographynophoto} {Nelio Cacho} (neliocacho@dimap.ufrn.br) is an associate professor in computer science at the Federal University of Rio Grande do Norte, Brazil. He has worked in the areas of software engineering and distributed systems for the last 15 years.
\end{IEEEbiographynophoto}

\vskip -2\baselineskip plus -1fil

\begin{IEEEbiographynophoto} {Claire-Isabelle Carlier} (claire-isabelle.carlier@brookfieldrenewable.com) is a business analyst with a background in cloud architecture and a data scientist with experience in the renewable energy sector. She is an Associate Member of IEEE.
\end{IEEEbiographynophoto}

\vskip -2\baselineskip plus -1fil

\begin{IEEEbiographynophoto} {Amy Corriveau} (corriveauam@cdmsmith.com) is the director of emerging business development for CDM Smith, Boston, Massachusetts, USA, a global engineering and consulting firm. She holds leadership positions within a number of industry associations including AWWA and WEF, leading various task forces, advisory committees, and technical committees.
\end{IEEEbiographynophoto}

\vskip -2\baselineskip plus -1fil

\begin{IEEEbiographynophoto} {Inas Khayal} (inas.khayal@dartmouth.edu) is an assistant professor in the Dartmouth Institute of Health Policy and Clinical Practice at the Geisel School of Medicine and adjunct assistant professor in the Department of Computer Science at Dartmouth College, Hanover, New Hampshire, USA. She is a Member of IEEE.
\end{IEEEbiographynophoto}

\vskip -2\baselineskip plus -1fil

\begin{IEEEbiographynophoto} {Barry Liner} (bliner@wef.org) is the chief technical officer at the Water Environment Federation (WEF), Alexandria, Virginia, USA, and is responsible for leading innovation and sustainability initiatives through WEF’s Water Science and Engineering Center, including founding the Innovation Pavilion at WEFTEC, the world’s largest annual water quality conference.
\end{IEEEbiographynophoto}

\vskip -2\baselineskip plus -1fil

\begin{IEEEbiographynophoto} {Joberto S.B. Martins} (joberto.martins@gmail.com) is a full professor at Salvador University, Brazil, and an international professor at Hochschule für Techknik und Wirtschaft des Saarlandes, Germany. He is a member of the IEEE Smart City committee, a former member of the IEEE Smart Grid Research committee, and an IEEE Senior Member.
\end{IEEEbiographynophoto}

\vskip -2\baselineskip plus -1fil

\begin{IEEEbiographynophoto} {Farokh Rahimi} is a senior vice president at Open Access Technology International, Inc., Minneapolis, Minnesota, USA. He is a Life Senior Member of IEEE and a member of the GridWise Architecture Council.
\end{IEEEbiographynophoto}

\vskip -2\baselineskip plus -1fil

\begin{IEEEbiographynophoto} {Rosaldo Rossetti} is a senior research fellow and member of the directive board of the Artificial Intelligence and Computer Science Lab as well as a faculty member with the executive committee of the Department of Informatics Engineering, University of Porto, Portugal. He is currently the chair of the IEEE Technical Activities Committee on Artificial Transportation Systems and Simulation.
\end{IEEEbiographynophoto}

\vskip -2\baselineskip plus -1fil

\begin{IEEEbiographynophoto} {Wester C.H. Schoonenberg} earned his B.Sc. degree in systems engineering and policy analysis management from the Delft University of Technology, The Netherlands, in 2014. After his bachelor’s degree, he started his M.Sc. degree studies at the Masdar Institute of Science and Technology. In the summer of 2015, he transferred with the LIINES to the Thayer School of Engineering at Dartmouth, Hanover, New Hampshire, USA, as a Ph.D. degree candidate, where he is working on interdependent smart city infrastructures.
\end{IEEEbiographynophoto}

\vskip -2\baselineskip plus -1fil

\begin{IEEEbiographynophoto} {Ashlynn Stillwell} is an associate professor and the Elaine F. and William J. Hall Excellence Faculty Scholar in Civil and Environmental Engineering at the University of Illinois at Urbana-Champaign, Illinois, USA. Her research interests include creating sustainable water and energy systems in a policy-relevant context, including projects on urban water and energy sustainability, water impacts of electric power generation, and green stormwater infrastructure.
\end{IEEEbiographynophoto}

\vskip -2\baselineskip plus -1fil

\begin{IEEEbiographynophoto} {Yinhai Wang} earned his master’s degree in computer science from the University of Washington (UW), Seattle, Washington, USA, and his Ph.D. degree in transportation engineering from the University of Tokyo, Japan, in 1998. He is a professor of transportation engineering and the founding director of the Smart Transportation Applications and Research Laboratory at UW.
\end{IEEEbiographynophoto}

\vskip -2\baselineskip plus -1fil

%\vfill

\end{document}